# Angular dependence of large negative magnetoresistance in a field-induced Weyl semimetal candidate HoAuSn

Yue Lu,[1,2] Jie Chen,[1,3*] Feng Zhou,[1,2] Yong-Chang Lau,[2,4] Piotr Wiśniewski,[5] Dariusz Kaczorowski,[5] Xuekui Xi,[2] and Wenhong Wang[1,2,3*]

[1]School of Electronics and Information Engineering, Tiangong University, Tianjin 300387, China

[2]Institute of Physics, Chinese Academy of Sciences, Beijing 100190, China

[3]Tianjin Key Laboratory of Optoelectronic Detection Technology and System, Tiangong University, Tianjin 300387, China

[4]University of Chinese Academy of Sciences, Beijing 100049, China

[5]Institute of Low Temperature and Structure Research, Polish Academy of Sciences, Okólna 2, Wrocław 50-422, Poland

**Abstract**

We have systematically studied the angular dependence of magnetoresistance (MR) in antiferromagnetic half-Heusler HoAuSn single crystals. Negative MR as large as ~99% is observed at 9 T, is not restricted to the specific configuration of applied magnetics fields and current, and can persist up to 20 K, much higher than the Néel temperature ($T_N$~1.9 K). First-principles calculations suggest that the observed large negative MR is derived from a field-induced transition from trivial semimetal to Weyl semimetal, and consequently, the spin scattering is significantly suppressed. Taking into consideration that large negative MR has so far been rarely reported, especially in antiferromagnetic materials, it is anticipated that the present work not only offers a guideline for searching materials with large negative MR but also helps to further realize other exotic topological electronic states in a large class of antiferromagnetic half-Heusler compounds.

**Email: chenjie@tiangong.edu.cn; wenhongwang@tiangong.edu.cn**

# 1 Introduction

Magnetoresistance (MR) describes the change in the resistivity of a material under the influence of an external magnetic field. The value of MR can be expressed as MR=[$\rho(B)$ -$\rho(0)$] /$\rho(0)$, where $\rho(B)$ and $\rho(0)$ are the resistivity of materials with and without the applied magnetic fields $B$, respectively. Over the past decades, the study of the magnetoresistance effect and its mechanism has been one of the research hotspots in condensed matter physics. For example, the giant magnetoresistance (GMR) phenomenon first discovered in Fe/Cr multilayers [1] has been widely used in spintronic devices such as hard disks [2, 3]. On the other hand, the in-depth study of anisotropic magnetoresistance (AMR) [4-8], tunneling magnetoresistance (TMR) [9, 10], and colossal magnetoresistance (CMR) [11] effects have greatly accelerated the pace of the information age. Among them, the negative MR effect, that is the resistivity decreases with the increase of the magnetic field, has unique advantages in terms of performance in magnetic sensors [12]. Therefore, it is of great practical value and research significance to search for materials with a large negative MR and elucidate the underlying mechanism.

In metals and semiconductors, a positive MR is common because electrons are affected by Lorentz forces. In contrast materials showing a large negative MR effect are relatively rare, and their MR usually comes from special physical mechanisms. For example, the Kondo effect [13], weak localization [14], field-induced paramagnetic semiconductor to ferromagnetic metal phase transition [15, 16]and longitudinal negative MR caused by chiral anomaly in topological materials [17, 18]. In recent years, it has been found that longitudinal negative MR can emerge in half-Heusler antiferromagnetic RPtBi (R is a lanthanide rare earth atom) and DyPdBi due to chiral anomaly [19-24]. In addition, large negative MR independent of the angle between the magnetic field and current has been reported recently in HoAuSn [25] and TbPdBi [26], where the field-induced transition from trivial semimetal to Weyl semimetal is responsible for the large negative MR.

In this paper, we have carried out detailed magnetotransport measurements on single crystals of antiferromagnetic half-Heusler HoAuSn. Due to the field-induced band structure change and the transition from semiconductor to semimetal, a Weyl point appears, resulting in a very large negative MR value of ~99%. Interestingly, such a large negative MR is very robust and can keep up to 20 K, much higher than its Néel temperature ($T_N$~1.9 K). Accompanied by the transition, the AMR effect also exhibit a significant change in symmetry.

## 2 Experimental

HoAuSn single crystals were grown by a self-flux method. Ho lumps, Au granules and Sn short granules with a molar ratio of 1:1:15 were used as starting materials and they were placed in an alumina crucible and vacuum sealed in a quartz tube. The ampoules were then placed in a chamber furnace and heated to 1000°C in 4 hours, maintained at that temperature for 24 hours, and subsequently cooled at 2 K/h to 500°C. The ampoules were quickly removed from the furnace and centrifuged. The crystal structure was characterized by a D/Max-2400 model X-ray diffractometer produced by Rigaku Company in Japan. Magnetoresistance was measured in a Physical Property Measurement System (PPMS), using the standard four-terminal method. Extremely low-temperature magnetic measurements were performed using the helium-3 option in a Quantum Design's (QD) Magnetic Property Measurement System (MPMS). Aberration-corrected scanning transmission electron microscopy (STEM) measurements were performed by JEOL ARM200F (JEOL, Tokyo, Japan) 200 keV transmission electron microscope. First-principle calculations were performed by VASP [27], using PBE-GGA [28] pseudopotential. Strong interatomic interactions in the 4*f* orbital of Ho were approximated by the GGA+U method, where $U_{eff}$ = 8 eV.

## 3 Results and discussion

Figures 1a and 1b show the perspective view and side view of a unit cell of HoAuSn crystallizing in half-Heusler structure *F*-43*m* (Space group No.216). Figure 1c shows the X-ray diffraction (XRD) spectrum of (111) plane of HoAuSn single crystal. From the peak positions, we extractd the lattice parameter *a*= 6.624 Å. The inset is the optical photograph of a single crystal with (111) plane. The crystal structure and the crystallographic axes were determined by single-crystal XRD, as shown in Figure S1. High-resolution STEM was also performed. Figure 1d shows the STEM image with atomic resolution of HoAuSn single crystal viewed along the normal of (110) plane. The periodic arrangement of atoms along the [001] crystal direction can be seen by local magnification and agrees with the crystallographic model. Similarly, the single crystal XRD pattern also shows that we have obtained high quality single crystals. Then we conducted energy-dispersive X-ray spectroscopy (EDX) analysis as shown in Figure S2. The distribution of elements is uniform and the atomic molar ratio is close to 1:1:1. Then we measured the thermal magnetization curve under $B$=0.05 T, as shown in Figures 1e and found the Néel transition point is 1.9 K. The Curie-Weiss fitting to the data collected in paramagnetic state indicates antiferromagnetic interactions. The effective magnetic moment of $Ho^{3+}$ is $10.7\mu_B$. Figure 1f shows the magnetization $M$ at different temperatures as a function of the magnetic field applied along [11-2]. The magnetization does not show a linear behavior at low temperature. At 0.4 K, below $T_N$ the $M$ shows a metamagnetic transition at 4 T and gradually approaching saturation.

Figure 2a shows the temperature-dependent resistivity curves recorded in different magnetic fields. The magnetic field is applied along the [1-10] direction, which is also parallel to the current. At 0 T, it shows a semiconductor-like behavior., that is the resistivity increases with decreasing temperature. As the magnetic field increase, the resistivity $\rho_{xx}$ at low temperature decreases rapidly and gradually converts to metallic behavior in the low temperature region. $\rho_{xx}$ drops by two orders of magnitude at $T$=2 K with a field of 9 T. Furthermore, we measured magnetoresistance with three different current and magnetic field orientation configurations, as shown in Figures 2b - 2d. The direction of the applied current remains along [1-10] direction all the time, while the magnetic field is applied to [111], [1-10], and [11-2] directions. At low temperatures, HoAuSn exhibits a positive magnetoresistance below $B < 1$ T. The MR rapidly turns to negative value as the magnetic field increases. The MR takes on an M-shape, which usually occurs in ferromagnetic materials due to the motion of the domain walls [29], but rarely occurs in antiferromagnetic materials. We infer that the positive magnetoresistance below 1 T may be caused by the Zeeman splitting of the 4$f$ state in the Ho atom, which increases the carrier scattering [25]. Accompanied by an increase in the magnetic field, spin polarization leads to degeneracy of the conduction and valence bands, which in turn leads to the closure of the band gap. Thus, the large negative MR dominates the MR curves. Below 20 K, the negative magnetoresistance is almost independent of temperature, but above 20 K, its magnitude decreases with increasing temperature. The thermal disturbance enhances the electron scattering and reduces the negative magnetoresistance. When the current is parallel to the magnetic field, MR gradually becomes saturated in magnetic field above 4 T. The negative magnetoresistance reaches 99%. However, for the other two cases where the magnetic field and current are orthogonal, it is different, and the magnetoresistance above 4 T shows a rising trend. At 9 T the negative magnetoresistance becomes 90%. Therefore, that the change in the angle between the magnetic field and the current has an obvious influence on the negative magnetoresistance.

In recent years, a lot of attention has been paid to materials with large negative MR. In the nonmagnetic Dirac semimetal $Cd_3As_2$[29], the giant negative MR 63% induced by the chiral anomaly. In the antiferromagnetic $BaMn_2Bi_2$[30], the strong Mott localization leads to the insulating behavior observed in the transport resistivity. when a magnetic field is applied perpendicular to the Néel vector of the AFM order, the insulating state transitions to the metallic state, resulting in negative MR up to 96%. In antiferromagnetic $EuMnSb_2$[31], the large negative MR is attributed to the metal-insulator transition that occurs when the lattice undergoes orthorhombic distortion. And in the half-heusler alloy TbPbBi[25], the high spin- polarization of carriers in half-topological semimetals leads to large negative MR. However, these materials are weakly anisotropy of MR. For the half-Heusler HoAuSn, we propose that the magnetic field changes the bandgap structure and the magnetic field induces the Weyl point to cause the chiral anomaly combined with a large negative MR. At the same time, there is also significant angular dependence of MR at high fields.

To understand the relationship between magnetoresistance and the angle between current and magnetic field, we have carried out detailed measurements of angular-dependent magnetoresistance. At 2 K, the current direction is fixed to [1-10]. The magnetic field rotates from [111] to [1-10], [1-10] to [11-2], and [111] to [11-2], respectively with magnetic field ranging from 0.4 T to 9 T, as shown in Figures 3a-3c. More detailed measurement data can be found in Figure S3. At 0.4 T, the same result was found for the three measurements, which was M-shaped AMR, that is the resistance is smaller when the field is transverse to the current. In this field range, the corresponding magnetoresistance is positive. At 0.8 T field, the curve is different. At this stage, the AMR changes dramatically, corresponding to the phase of rapid decline in magnetoresistance. At the 9 T field, when the magnetoresistance is saturated, the AMR changes completely with a W-shaped angular dependence. To understand the cause of complex AMR, we fit it with terms having two-fold and four-fold symmetries. The two-fold symmetry may come from chiral anomalies, while the four-fold symmetry may reflect the crystal or magnetic symmetry of the corresponding plane. Similarly, we also measured the angle dependent AMR with current along the (001) plane and obtained a clearer AMR signal with two-fold and four-fold symmetry in the weak and strong field range, respectively. Details are shown in Figure S4. The AMR changes with the different fields, indicating that the magnetic field changes may have a significant effect on the band structure.

According to the experimental results, we believe that the large negative magnetoresistance in HoAuSn is related to the change of band structure induced by the magnetic field [33, 34]. To provide more insights into the origin of surprising sensitivity of HoAuSn to external magnetic field, we have calculated the detailed band structure of HoAuSn and its Fermi surface using first-principles calculations, all of which take into account the spin-orbit coupling (SOC). Figure 4a shows the band structures of the non-magnetic (NM) state, which suggests that HoAuSn is a semimetal with a small hole pocket near Γ point and an electron pocket near X point. Moreover, in order to understand the band structure after adding the magnetic field, we have calculated the band structures of ferromagnetic (FM) state, where *M* along the [111], as shown in Figure 4c. The band structures of the FM state is similar to that of the NM state, but Weyl points appear in the Γ-L path. This is one of the reasons for the large transverse negative magnetoresistance and two-fold symmetry of AMR. In addition, we have calculated the band structures of FM state, where *M* along the [110] and [001]. It has been demonstrated through Wilson-loop calculations that Weyl points exist near the Fermi level in both polarization directions symmetrically, as shown in Figure S5 and Figure S6. We calculated the density of states and the results are shown in Figure 4a and 4c. We find that the FM state has a higher density of states than the NM state at the Fermi surface, indicating that HoAuSn has more energy bands passing through the Fermi level at the FM state. Interestingly, it can also be intuitively seen that the area of Fermi surface of FM is larger than NM, as shown in Figures 4b and 4d. On the other hand, spin polarization in FM results in band splitting, which leads to the increase of the Fermi surface cross section in the FM state. It can be explained that the modification of the energy band by the magnetic field leads to negative magnetoresistance and the change of AMR symmetry.

## 4 Conclusion

In conclusion, we report that in the antiferromagnetic half-Heusler compound HoAuSn, the magnetic field changes the bandgap structure and induces the Weyl point to cause the chiral anomaly combined with a large negative magnetoresistance. We measured the resistivity for different angles between the current and magnetic field. In addition to negative MR, we observe by first principles calculations that HoAuSn is a topologically trivial semimetal in the NM state, and then we find Weyl points in the Γ-L path in the FM state. The reason for the large negative magnetoresistance in HoAuSn is explained, and one can expect to further regulate the negative magnetoresistance by replacing Ho by other lanthanide rare earth elements.

**Acknowledgements**

This work was supported by the National Key R&D Program of China (Grant no. 2022YFA1402600), and the National Natural Science Foundation of China (Grants Nos. 12304150 and 52161135108). A portion of this work was carried out at the Synergetic Extreme Condition User Facility (SECUF).Also, the authors acknowledge the support by the National Science Centre (Poland), grant 2021/40/Q/ST5/00066.

# Figure Captions:

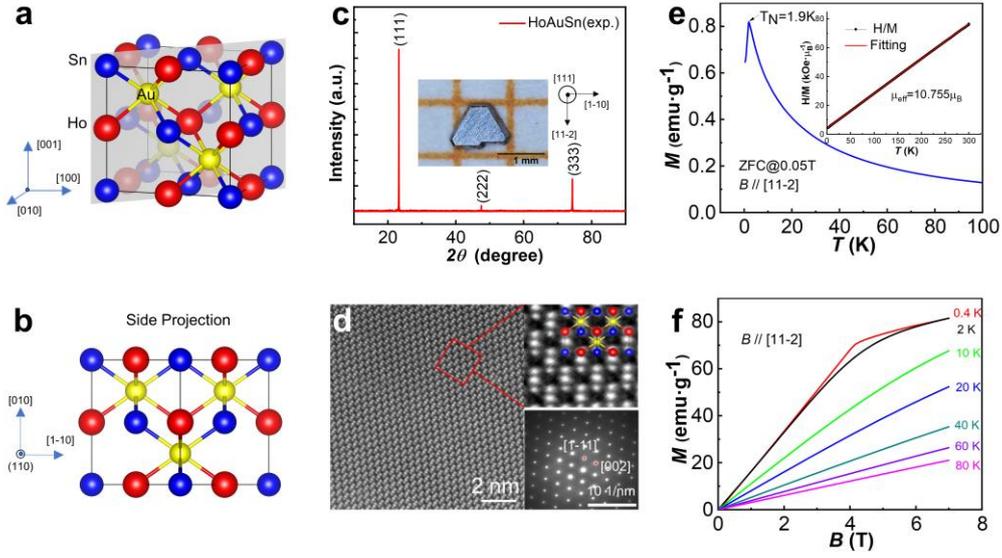

**FIG.1. a** Crystal structure of HoAuSn (space group *F*-43*m*). Ho, Au and Sn atoms are false colored to appear red, yellow, and blue, respectively. **b** Side view of HoAuSn unit cell along [110] direction. **c** XRD patterns of HoAuSn single crystal. The reflected (111) peaks demonstrate the orientation of the single crystal. Inset: Photographs of HoAuSn single crystals with the scalebar of 1 mm, and the crystal orientation marked. **d** Atomic configuration of the (110) plane detected by high angle annular dark field (HAADF) imaging. Inset: Local amplification and single crystal electron diffraction patterns. **e** Zero-field cooling (ZFC) curve of HoAuSn single crystal under an external magnetic field of *B*=0.05T along the [11-2] direction. The Néel temperature $T_N$=1.9 K (marked with a black arrow). The inset shows the fitting of a thermomagnetic curve with the Curie-Weiss law. **f** Magnetization curves with *B* along the [11-2] direction recorded at several temperatures.

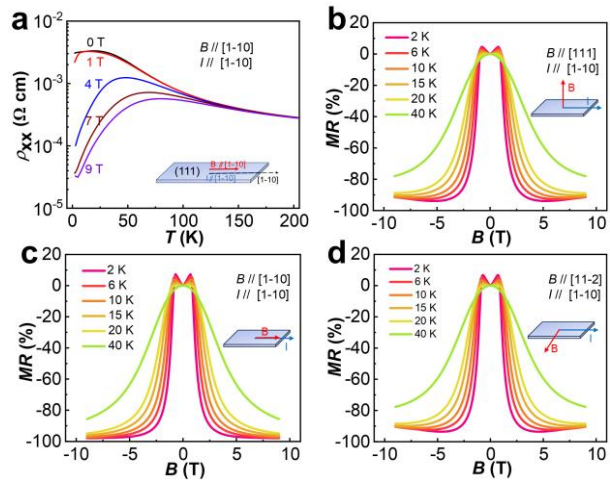

**FIG.2. a** Temperature dependence of the $\rho_{xx}$ under different magnetic fields for HoAuSn. **b-d** Magnetic field dependence of MR at various temperatures for different magnetic fields and current angle configurations.

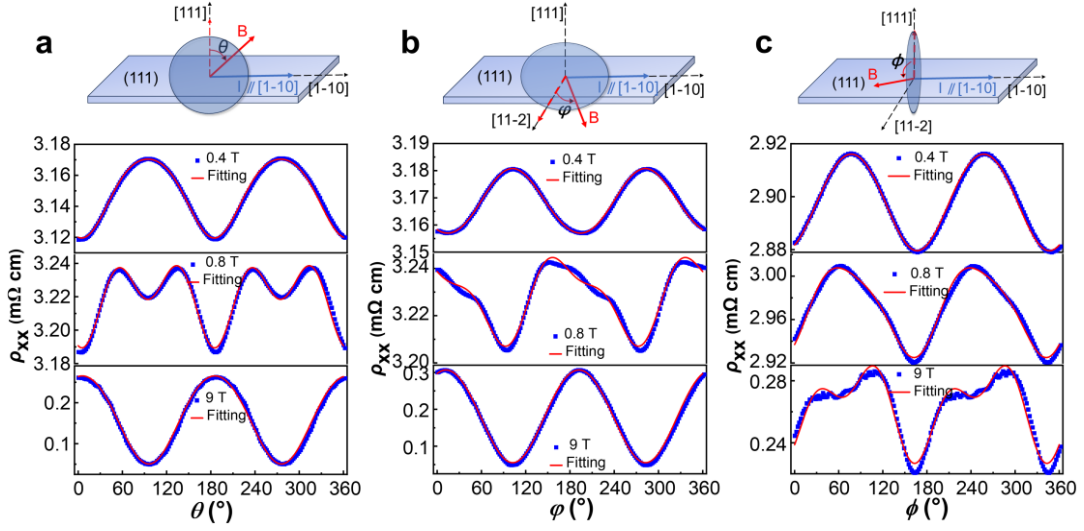

**FIG. 3.** Angular dependence of $\rho_{xx}$ at $T$=2K, in fields of 0.4, 0.8 and 9 T, with current along the direction [1-10], for different magnetic field rotation modes sketched in the uppermost row. **a** The magnetic field rotates from [111] to [1-10], and $\theta$ is defined as the angle between [111] and the magnetic field. **b** The magnetic field rotates from [1-10] to [11-2], and $\varphi$ is defined as the angle between [1-10] and the magnetic field. **c** The magnetic field rotates from [111] to [11-2], and $\phi$ is defined as the angle between [111] and the magnetic field.

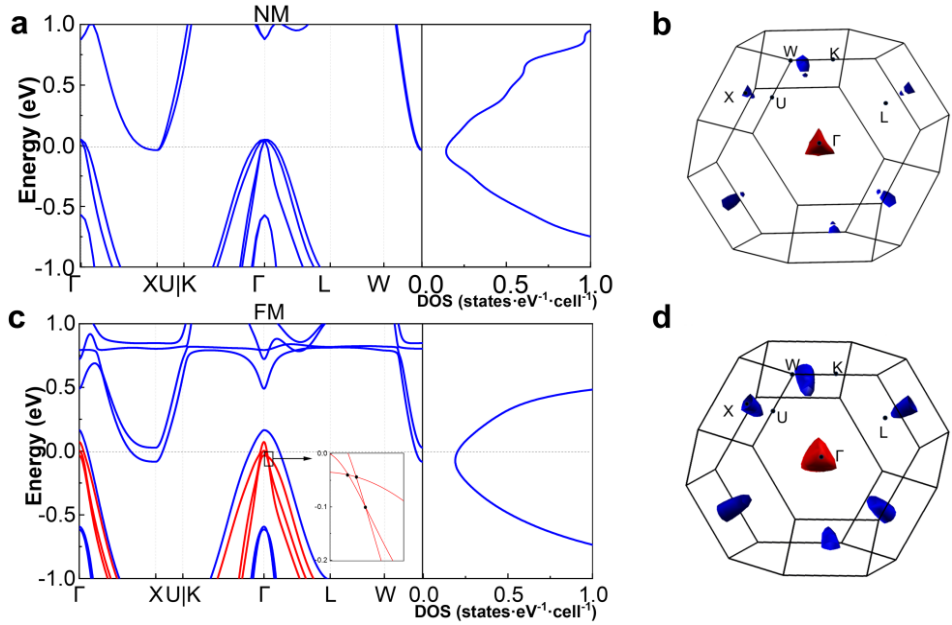

**FIG. 4.** HoAuSn from GGA+SOC+U ($U_{eff}$=8 eV) calculation **a** Band structures and density of states for the paramagnetic state. **b** The shape of the Fermi surfaces of HoAuSn in the first Brillouin zone derived from the band structure. **c** Band structures and density of states for ferromagnetic state. **d** Calculation Fermi surface for FM state.

**References**

[1] Baibich M. N., Broto J. M., Fert A., Van Dau F. N., Petroff F., Etienne P., Creuzet G., Friederich A. and Chazelas J., Giant Magnetoresistance of (001)Fe/(001)Cr Magnetic Superlattices, Physical Review Letters, 1988, 61(21): 2472.

[2] Fert A., Nobel Lecture: Origin, development, and future of spintronics, Reviews of Modern Physics, 2008, 80(4): 1517.

[3] Chappert C., Fert A. and Van Dau F. N., The emergence of spin electronics in data storage, Nature Materials, 2007, 6(11): 813.

[4] Dai Y., Zhao Y. W., Ma L., Tang M., Qiu X. P., Liu Y., Yuan Z. and Zhou S. M., Fourfold Anisotropic Magnetoresistance of L10 FePt Due to Relaxation Time Anisotropy, Physical Review Letters, 2022, 128(24): 247202.

[5] Oogane M., McFadden A. P., Kota Y., Brown-Heft T. L., Tsunoda M., Ando Y. and Palmstrøm C. J., Fourfold symmetric anisotropic magnetoresistance in half-metallic $Co_2MnSi$ Heusler alloy thin films, Japanese Journal of Applied Physics, 2018, 57(6): 063001.

[6] Liu Q.-Q., Yang G., Zhang J.-Y., Feng G.-N., Feng C., Zhan Q., Li M.-H. and Yu G.-H., Tunable perpendicular anisotropic magnetoresistance in CoO/Co/Pt heterostructures, Rare Metals, 2017, 42(2): 579.

[7] Liu J.-J., Meng K.-K., Chen J.-K., Wu Y., Miao J., Xu X.-G. and Jiang Y., Robust interface-induced unusual anomalous Hall effect in $Mn_3Sn$/Pt

bilayers, Rare Metals, 2022, 41(9): 3012.

[8] Vlietstra N., Shan J., Castel V., van Wees B. J. and Ben Youssef J., Spin-Hall magnetoresistance in platinum on yttrium iron garnet: Dependence on platinum thickness and in-plane/out-of-plane magnetization, Physical Review B, 2013, 87(18): 184421.

[9] Song T. C., Cai X. H., Tu M. W. Y., Zhang X. O., Huang B. V., Wilson N. P., Seyler K. L., Zhu L., Taniguchi T., Watanabe K., McGuire M. A., Cobden D. H., Xiao D., Yao W. and Xu X. D., Giant tunneling magnetoresistance in spin-filter van der Waals heterostructures, Science, 2018, 360(6394): 1214.

[10] Parkin S. S. P., Kaiser C., Panchula A., Rice P. M., Hughes B., Samant M. and Yang S.-H., Giant tunnelling magnetoresistance at room temperature with MgO (100) tunnel barriers, Nature Materials, 2004, 3(12): 862.

[11] Ramirez A. P., Colossal magnetoresistance, Journal of Physics-Condensed Matter, 1997, 9(39): 8171.

[12] Alekseev P. S., Negative Magnetoresistance in Viscous Flow of Two-Dimensional Electrons, Physical Review Letters, 2016, 117(16): 166601.

[13] Kondo J., RESISTANCE MINIMUM IN DILUTE MAGNETIC ALLOYS, Progress of Theoretical Physics, 1964, 32(1): 37.

[14] Altshuler B. L., Khmel'nitzkii D., Larkin A. I. and Lee P. A., Magnetoresistance and Hall effect in a disordered two-dimensional

electron gas, Physical Review B, 1980, 22(11): 5142.

[15] Seo J., De C., Ha H., Lee J. E., Park S., Park J., Skourski Y., Choi E. S., Kim B., Cho G. Y., Yeom H. W., Cheong S.-W., Kim J. H., Yang B.-J., Kim K. and Kim J. S., Colossal angular magnetoresistance in ferrimagnetic nodal-line semiconductors, Nature, 2021, 599(7886): 576.

[16] Yin J., Wu C., Li L., Yu J., Sun H., Shen B., Frandsen B. A., Yao D.-X. and Wang M., Large negative magnetoresistance in the antiferromagnetic rare-earth dichalcogenide $EuTe_2$, Physical Review Materials, 2020, 4(1): 013405.

[17] Liu E., Sun Y., Kumar N., Muechler L., Sun A., Jiao L., Yang S.-Y., Liu D., Liang A., Xu Q., Kroder J., Süß V., Borrmann H., Shekhar C., Wang Z., Xi C., Wang W., Schnelle W., Wirth S., Chen Y., Goennenwein S. T. B. and Felser C., Giant anomalous Hall effect in a ferromagnetic kagome-lattice semimetal, Nature Physics, 2018, 14(11): 1125.

[18] Wang Y., Liu E., Liu H., Pan Y., Zhang L., Zeng J., Fu Y., Wang M., Xu K., Huang Z., Wang Z., Lu H.-Z., Xing D., Wang B., Wan X. and Miao F., Gate-tunable negative longitudinal magnetoresistance in the predicted type-II Weyl semimetal $WTe_2$, Nature Communications, 2016, 7(1): 13142.

[19] Chen J., Li H., Ding B., Liu E., Yao Y., Wu G. and Wang W., Chiral-anomaly induced large negative magnetoresistance and nontrivial π-Berry phase in half-Heusler compounds RPtBi (R=Tb, Ho, and Er), Applied Physics Letters, 2020, 116(22): 222403.

[20] Hirschberger M., Kushwaha S., Wang Z., Gibson Q., Liang S., Belvin Carina A., Bernevig B. A., Cava R. J. and Ong N. P., The chiral anomaly and thermopower of Weyl fermions in the half-Heusler GdPtBi, Nature Materials, 2016, 15(11): 1161.

[21] Shekhar C., Kumar N., Grinenko V., Singh S., Sarkar R., Luetkens H., Wu S.-C., Zhang Y., Komarek A. C., Kampert E., Skourski Y., Wosnitza J., Schnelle W., McCollam A., Zeitler U., Kübler J., Yan B., Klauss H. H., Parkin S. S. P. and Felser C., Anomalous Hall effect in Weyl semimetal half-Heusler compounds RPtBi (R = Gd and Nd), Proceedings of the National Academy of Sciences, 2018, 115(37): 9140.

[22] Suzuki T., Chisnell R., Devarakonda A., Liu Y. T., Feng W., Xiao D., Lynn J. W. and Checkelsky J. G., Large anomalous Hall effect in a half-Heusler antiferromagnet, Nature Physics, 2016, 12(12): 1119.

[23] Guo C. Y., Wu F., Wu Z. Z., Smidman M., Cao C., Bostwick A., Jozwiak C., Rotenberg E., Liu Y., Steglich F. and Yuan H. Q., Evidence for Weyl fermions in a canonical heavy-fermion semimetal YbPtBi, Nature Communications, 2018, 9(1): 4622.

[24] Pavlosiuk O., Kaczorowski D. and Wiśniewski P., Negative longitudinal magnetoresistance as a sign of a possible chiral magnetic anomaly in the half-Heusler antiferromagnet DyPdBi, Physical Review B, 2019, 99(12): 125142.

[25] Ueda K., Yu T., Hirayama M., Kurokawa R., Nakajima T., Saito H., Kriener M., Hoshino M., Hashizume D., Arima T.-h., Arita R. and Tokura Y., Colossal negative magnetoresistance in field-induced Weyl semimetal of magnetic half-Heusler compound, Nature Communications, 2023, 14(1): 6339.

[26] Zhu Y., Huang C.-Y., Wang Y., Graf D., Lin H., Lee S. H., Singleton J., Min L., Palmstrom J. C., Bansil A., Singh B. and Mao Z., Large anomalous Hall effect and negative magnetoresistance in half-topological semimetals, Communications Physics, 2023, 6(1): 346.

[27] Kresse G. and Furthmuller J., Efficient iterative schemes for ab initio total-energy calculations using a plane-wave basis set, Physical Review B, 1996, 54(16): 11169.

[28] Perdew J. P., Burke K. and Ernzerhof M., Generalized gradient approximation made simple, Physical Review Letters, 1997, 78(7): 3865.

[29] Tang J.-X., Wang P.-H., You Y.-R., Wang Y.-D., Xu Z., Hou Z.-P., Zhang H.-G., Xu G.-Z. and Xu F., Abnormal low-field M-type magnetoresistance in hexagonal noncollinear ferromagnetic MnFeGe alloy, Rare Metals, 2022, 41(8): 2680.

[30] Li C.-Z., Wang L.-X., Liu H., Wang J., Liao Z.-M. and Yu D.-P., Giant negative magnetoresistance induced by the chiral anomaly in individual $Cd_3As_2$ nanowires, Nature Communications, 2015, 6(1): 10137.

[31] Ogasawara T., Huynh K.-K., Tahara T., Kida T., Hagiwara M., Arčon

D., Kimata M., Matsushita S. Y., Nagata K. and Tanigaki K., Large negative magnetoresistance in the antiferromagnet BaMn$_2$Bi$_2$, Physical Review B, 2021, 103(12): 125108.

[32] Yi C., Yang S., Yang M., Wang L., Matsushita Y., Miao S., Jiao Y., Cheng J., Li Y., Yamaura K., Shi Y. and Luo J., Large negative magnetoresistance of a nearly Dirac material: Layered antimonide EuMnSb$_2$, Physical Review B, 2017, 96(20): 205103.

[33] Felser C. and Yan B., Magnetically induced, Nature Materials, 2016, 15(11): 1149.

[34] Cano J., Bradlyn B., Wang Z., Hirschberger M., Ong N. P. and Bernevig B. A., Chiral anomaly factory: Creating Weyl fermions with a magnetic field, Physical Review B, 2017, 95(16): 161306.